\documentclass[pra,aps,twocolumn,byrevtex,showpacs]{revtex4}
\usepackage{epsfig}
\begin{document}

\title{Experimental Investigation of the Robustness of Partially
Entangled Photons over 11km}
\author{R.~T.~Thew$^{1}$ \thanks{Email: Robert.Thew@physics.unige.ch}}
\author{S.~Tanzilli$^{1,2}$}
\author{W.~Tittel$^{1,3}$}
\author{H.~Zbinden$^{1}$}
\author{N.~Gisin$^{1}$}

\affiliation{$^{1}$Group of Applied Physics, University of Geneva,
1211 Geneva 4, Switzerland}

\affiliation{$^{2}$ Laboratoire de Physique de la Mati\`ere
Condens\'ee, CNRS UMR 6622, Universit\'e de Nice-Sophia Antipolis,
Parc Valrose, 06108 Nice Cedex 2, France}

\affiliation{$^{3}$Danish Quantum Optics Center, Institute for
Physics and Astronomy, University of Aarhus, Denmark}

\date{\today}

\begin{abstract}
We experimentally investigate the robustness of maximal and
non-maximal Time-Bin entangled photons over distances up to
11\,km. The entanglement is determined by controllable parameters
and in all cases is shown to be robust, in that the photons
maintain their degree of entanglement after transmission.
\end{abstract}
 \maketitle

Quantum communication and quantum networks are only part of a much
larger field now of Quantum Information Science \cite{Nielsen00a}.
At the heart of many of these associated endeavours is
entanglement \cite{Tittel01a}. Entangled states also play an
essential role with respect to the fundamental nature of the
microscopic world as investigated in tests of Bell inequalities
\cite{Einstein35, Bell64, Aspect82, Tittel98, Weihs98}. Beyond the
mere existence of entanglement, quantum communication schemes like
quantum cryptography \cite{Eckert91, Gisin02a} and teleportation
\cite{Bennett93a, Bouwmeester97a, Rome98a, Shih01a} have been
developed to utilise what has become known as this {\it quantum
resource}, entanglement. Schemes using photonic quantum channels
that connect distant nodes of a quantum network \cite{Cirac97a} or
for quantum repeaters \cite{Breigel98a} rely on the possibility to
broadcast entanglement over significant distances.

An important aspect which has received little attention from an
experimental perspective is that information may need to be
encoded in possibly unknown states of arbitrary degrees of
entanglement. At a fundamental level, we expect that the
entanglement should be robust whether the state is maximally
entangled or not. Depending on the decohering environment this may
not be the case. At some intuitive level however, we may be
inclined to believe that due to their inherent symmetry, maximally
entangled states are more robust against decoherence than
non-maximal ones. We will consider a state to be robust if, when
it is detected, the initial entanglement is unchanged over the
length of the transmission.

We have previously shown that maximally entangled Energy-Time
photons are robust enough to violate a Bell inequality between
analysers 10\,km apart \cite{Tittel98}. In this letter, we now
investigate the robustness of partially, and maximally, entangled
Time-Bin {\it Qubits}, when transmitted over long optical fibers,
along the way providing the necessary basis for distribution of
arbitrary states.

\begin{figure}
\epsfig{figure=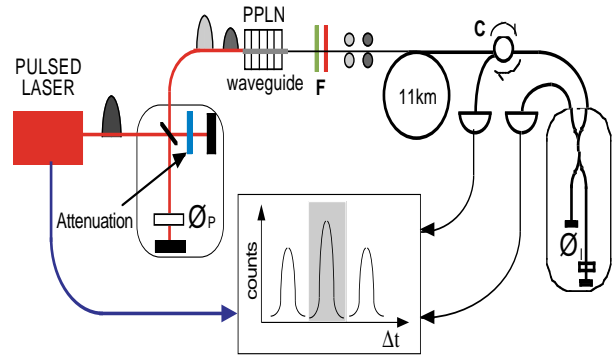,width=80mm,height=50mm}
\caption{Experimental schematic: A pulsed laser diode source and
Michelson interferometer produce 2 pump pulses which are then
incident on a PPLN waveguide producing 2 entangled photons. After
filtering\,(F), each pair is collected and transmitted along a
fibre spool to a fibre Michelson interferometer. A circulator\,(C)
allows input and detection on the same port. A triple coincidence
between 2 photons and one of the pump pulses then detects the
entangled state.\label{fig:expschem} }
\end{figure}

Entangled time-bin photons\,(entangled qubits) \cite{Brendel99a}
can be created using the experimental setup pictured in
Fig.(\ref{fig:expschem}). A coherent superposition of two
classical pump pulses is generated, from a single diode laser
pulse, after passing through a Bulk optics Michelson
interferometer with a large path-length difference. The laser
produces pulses of less than 100\,ps width at a frequency of
80\,MHz with a wavelength of 657\,nm. In this scheme the pulse
duration must be short compared to the travel time difference of
the pump interferometer, 1.2\,ns in our case. The two pulses then
undergo spontaneous parametric downconversion in a Periodically
Poled Lithium-Niobate\,(PPLN) waveguide \cite{Tanzilli01a,
Tanzilli01b} producing pairs of entangled photon at 1314\,nm
wavelength, convenient for fibre telecommunication. At these
wavelengths the detection is obtained via passively quenched
germanium avalanche photodiodes, operated in Geiger mode and
cooled to 77\,K. Depending on the amplitudes of the two classical
pulses and the relative phase, one can create maximally and
non-maximally entangled states of the form
\begin{equation}
|\psi\rangle =  \alpha |1,0;1,0\rangle_{AB} + \beta e^{i\phi_{P}}
|0,1;0,1\rangle_{AB}\label{eq:time-bin}
\end{equation}
where $\alpha$ and $\beta$ are real and $\alpha^2 + \beta^2 = 1$,
due to normalisation.
 Our notation represents, in the case of a state like $|n,0;0,m
\rangle$, that $n$ photons are in the first time-bin for $A$ and
that $m$ photons are in the second time-bin for $B$. The time
difference is obtained as a result of the photons having taken
either the short or long arms of the Bulk, or "Pump",
interferometer, and  $\phi_P$ is the resulting relative phase. For
equal amplitudes, $\alpha=\beta$, the state is maximally
entangled, and when $\phi_{P}$=0, Eq.(\ref{eq:time-bin})
corresponds to the maximally entangled Bell state
$|\phi^+\rangle$.

After passing through the second\,(fibre) interferometer the state
can be described by:
\begin{eqnarray}
|\psi'\rangle =  \alpha |1,0,0;1,0,0\rangle_{AB} + \alpha
e^{2i\phi_{I}} |0,1,0;0,1,0\rangle_{AB} \nonumber\\  + \beta
e^{i\phi_{P}} |0,1,0;0,1,0\rangle_{AB} + \beta
e^{i(2\phi_{I}-\phi_{P})} |0,0,1;0,0,1\rangle_{AB}
\label{eq:time-bin-2int}.
\end{eqnarray}
The resulting double coincidences for the pump and A\,(and the
pump and B) from this state correspond to the three peaks of the
coincidence histogram, as depicted at the bottom of
Fig.(\ref{fig:expschem}). The two middle terms of this state
interfere with respect to the amplitudes and phases of our initial
entangled state in the central time bin. We can distinguish the
first and last terms in Eq.(\ref{eq:time-bin-2int}) via the pump
timing information. Thus, conditioning the detection on events in
both middle peaks by making a triple coincidence corresponds to a
projective measurement onto the state of Eq.(\ref{eq:time-bin}).

In this experiment we are not concerned with questions of
non-locality but simply with the robustness of the Time-Bin
entangled states. As such we primarily utilise a "Franson
Repli\'e" arrangement which utilises only one analyser
interferometer as depicted in Fig.(\ref{fig:expschem}). This
arrangement can be thought of as having used the symmetry of the
standard Franson interferometer setup \cite{Franson89a}  and
folded it in half so both interferometers were on top of each
other in a sense. After the 11\,km of fibre\,(on a spool) a
circulator is placed at one port of the interferometer allowing us
to both input and detect. The other detector operates normally on
the other port. Coincidences correspond to both photons taking the
short or the long paths together as opposed to doing it in
independent interferometers. Whilst we believe that the
correlations will be the same, whether we use one or two
interferometers, we will give some results with respect to an
experimental arrangement with two interferometers so that each
photon travels over an independent 2.4\,km length of fibre. This
is done by placing a 2x2 fibre coupler\,(beamsplitter) after the
PPLN waveguide in Fig.(\ref{fig:expschem}), connected to the two
fibres, each with an interferometer, where detection is made on
one output of each.

During transmission these pure states can in general suffer from
environment induced phase as well as from bit flips. Bit flips can
happen if the broadening of the photon\,(in time space), due to
chromatic dispersion effects in fibers, is such that the amplitude
initially located in one time bin starts overlapping with the one
in the other time bin. ie the three peaks at the bottom of
Fig.(\ref{fig:expschem}) merge together. This reduces our ability
to discriminate between them, and hence we loose information. This
can be prevented in several ways: we already produce our entangled
photons centered at telecommunication wavelength of 1.3\,$\mu$m
where chromatic dispersion is zero, thus reducing the
susceptibility to dispersion. We can improve this non-locally by
using the right choice of wavelength determined by the dispersion
curve of the fibers, in the same way that it is done for
energy-time entangled states \cite{Franson92a, Tittel99a}.
However, note that the compensation will be less perfect as the
energy correlation between the entangled qubits is less stringent.
Further on, we can utilise interference filters, in our case of
40\,nm width, to reduce the spectral width of the photons, leading
to a less pronounced spread of the pulse during a given
transmission distance. And finally, we can increase the separation
between the time bins, but this would make the interferometers
larger and less stable and render the qubits more vulnerable to
phase errors during transmission.

Phase errors generally arise as a result of some random variation
between the two time bins but this is negligible in our case, at
least as far as  transmission is concerned. With 1.2\,ns
separation between time bins there would need to be very
high\,(GHz) frequency noise in order to produce a phase flip. Our
main phase error arises during creation and detection of ensembles
of photons over extended periods of time. This requires that the
interferometers have the same path-length difference  for the
entire time that we prepare and measure a particular state. This
will be discussed further as we discuss our states.

If we are going to generate non-maximally entangled states, then
an important question is: how can we characterise them? Firstly we
can restrict our attention to pure states of the form of
Eq.(\ref{eq:time-bin}), as we post-select the final state. Now,
given the state in Eq.(\ref{eq:time-bin-2int}), the probability of
coincidence in the central time-bin is given by,
\begin{eqnarray}
P_c =0.5[ \alpha^2 + \beta^2 + 2\alpha \beta
cos(\phi)].\label{eq:probcoinc}
\end{eqnarray}
We see that the probability of a coincidence detection varies with
the phase, $\phi = 2\phi_I -\phi_P$. Therefore, by varying the
phase, we can scan through maximum and minimum probabilities
corresponding to regimes of maximum constructive and destructive
interference and thus determine the Visibility which is given by,
\begin{eqnarray}
 V = 2\alpha \beta.\label{eq:vis}
\end{eqnarray}
Our fibre interferometers are thermally stabilised and the phase
is varied by changing the temperature of the interferometer.
\begin{figure}[h]
\epsfig{figure=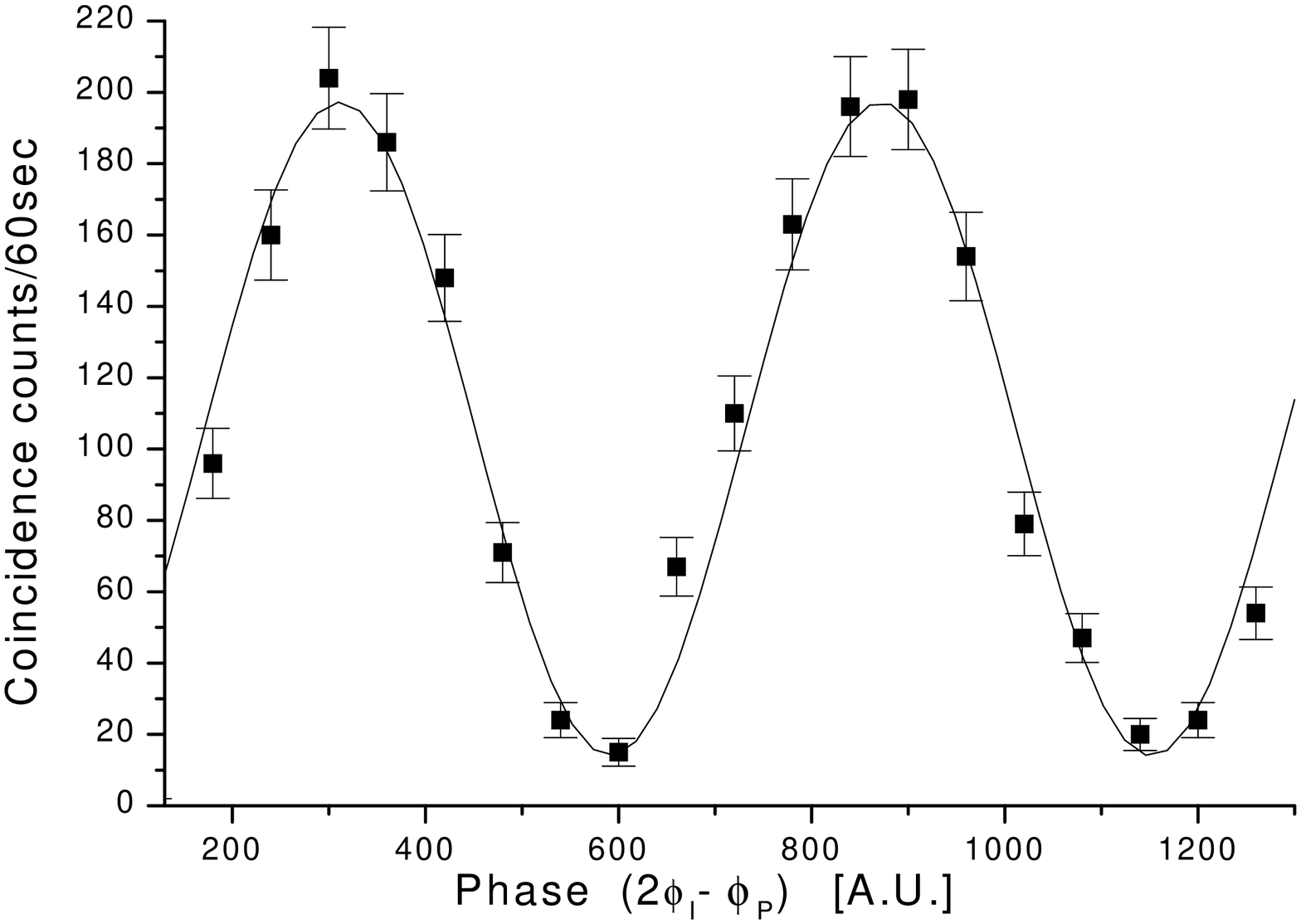,width=80mm,height=48mm} \caption{The nett
Interference fringes for the experimental setup shown in
Fig.(\ref{fig:expschem}) after the photons had travelled 11\,km.
\label{fig:ldnmvis} }
\end{figure}

In Fig.({\ref{fig:ldnmvis}) we see the results of this approach
using the experimental setup in Fig.(\ref{fig:expschem}) with
11\,km of fibre. Depicted are the nett count rates after
subtraction of accidental coincidences. We will comment on this
later. Due to the long distances the signal is reduced and hence
we have to count for a longer time, 60 seconds in this instance.
The squares denote the number of triple coincidences in each 60
second interval. It is for this 60 second integration time that we
have to maintain the stability of the laser to within a fraction
of a wavelength to minimise the phase errors discussed previously.
The solid line that traces these points is a sinusoidal fit to the
data derived from Eq.(\ref{eq:probcoinc}) and allows us to
determine the Visibility.

The key here is that the nett Visibility parameterises the
entanglement in the final state, as we would expect. Both the
bit-flips and phase flips will manifest themselves as a reduction
in Visibility and as such we will use this as the experimental
measure of our entanglement. With respect to
Fig.({\ref{fig:ldnmvis}), this analysis returns a value of $V =
94.2 \pm 4.8\%$ after transmitting the maximally entangled state
over 11km. The error in this result, $\Delta V$, is determined by
the numerical uncertainty in the sinusoidal fit to the data.

Table(\ref{table:vis}) shows the raw Visibility as well as the
nett Visibility, which is derived after subtracting the accidental
coincidences, for maximally entangled states with both one and two
interferometers. We see that although the raw Visibility is less
over the longer distances, the nett Visibilities are almost
equivalent.
\begin{table}[h]
\begin{tabular}{ccccc}
\hline \hline \hspace{4mm}Run \hspace{4mm} & \hspace{4mm}Distance
\hspace{4mm}& \hspace{4mm}$V_{raw}\hspace{4mm}$ &
\hspace{4mm}$V_{nett}\hspace{4mm}$ & \hspace{4mm}$\Delta V
\hspace{4mm}$
\\ \hline\hline

I & 0\,km  &  90.2 & 94.9 & 3.7\\ \hline

II & 11\,km & 86.8 & 94.2 & 4.8 \\ \hline \hline

III & 0\,km  & 89.1 & 92.7 & 4.7 \\\hline

IV & 2.4\,km & 84.5 & 92.2  & 4.8 \\ \hline
\end{tabular}
\caption{The raw and nett\,(accidental coincidences subtracted)
Visibilities for the maximally entangled states. I and II use 1
interferometer and rows III and IV use 2. $\Delta V$ is the
uncertainty in the Visibility.} \label{table:vis}
\end{table}

 Let us briefly comment on the nature of accidental
coincidences. They occur if both events in the coincidence window
are triggered by noise. Also, there will be contributions if one
photon triggers one detector and noise triggers the other while
the photon's correlated partner is absorbed in the fiber. Thus,
the resulting decrease of raw visibility is due to a combination
of fiber losses and detector noise. However, since we are
interested in the impact of bit and phase flips on the
entanglement, we must subtract this source of noise which is well
understood and is easily measured.

 The other aspect of characterisation is related to
generating the non-maximally entangled states. We can quantify the
expected entanglement in the state in terms of the Entropy of
Entanglement:
\begin{eqnarray}
E = - \alpha^2log_2 \alpha^2 -  (1- \alpha^2)log_2(1-
\alpha^2)\label{eq:entangle},
\end{eqnarray}
where we use the amplitudes of the two pump pulses, appropriately
normalised. With this experimental setup non-maximally entangled
states can be realised in a controlled manner by varying the
attenuation in one of the arms of the pump interferometer so that
the classical pulses have unequal magnitudes. In practice we vary
the attenuation in both arms so that we have the same mean power
before the waveguide.
\begin{figure}[h]
\epsfig{figure=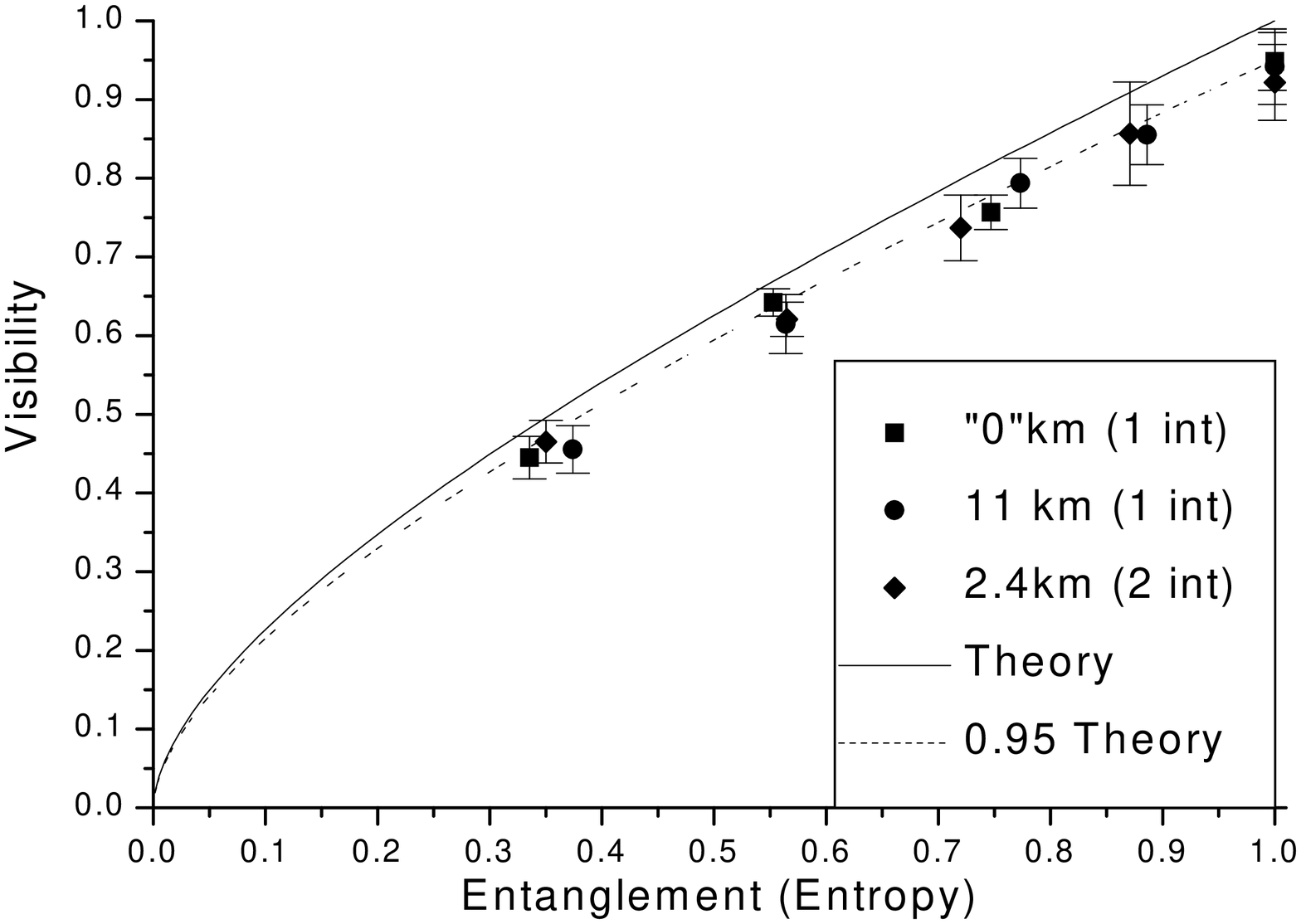,width=80mm,height=49mm}
  \caption{Nett Visibility of the output state as a function of the expected
Entanglement, measured at our source. The relationship is
unchanged by transmission over significant distances.
\label{fig:ldVvsE} }
\end{figure}

The results of the measurements of the non-maximally entangled
states are given in Fig.(\ref{fig:ldVvsE}). The solid line shows
the theoretically expected Visibility from Eq.(\ref{eq:vis}) as a
function of the Entanglement Eq.(\ref{eq:entangle}) as $\alpha$
and $\beta$ are varied. In the experiment we have complete control
over the classical amplitudes and these are measured directly at
the output of the pump interferometer. After transmission the nett
Visibility are obtained from sinusoidal fits described previously.
On average, subtraction of the noise improved the raw Visibilities
for the 0 distance runs by less than 5\% and the 11\,km runs by
less than 9\%. A dashed line corresponding to the theory but
scaled to have a maximum Visibility of 95\% is also shown. We see
that for both the experimental setup as depicted in
Fig.(\ref{fig:expschem}), with or without 11\,km of fibre, and
also with one or with two interferometers the results are in good
agreement with the theory. Specifically we see that regardless of
the initial entanglement or the distance travelled that there is
no loss of entanglement over transmission for any of the states.

With our experimental configuration we don't have access to an
absolute phase reference within the system and hence standard
tomographic techniques \cite{White99a,James01a,Thew02a} won't
work. Given that the entanglement could be parameterised by the
Visibility and there were no adverse effects due to decoherence,
there appears little to be gained by reconstructing the full
density matrix for the state.

In this experiment we use a PPLN waveguide which provides a highly
efficient generator of entangled photons. A consequence of this
high efficiency, conversion rates 4 orders of magnitude higher
than obtained with bulk sources \cite{Tanzilli01a, Tanzilli01b},
is that when using a pulsed laser the probability of producing
multiple photon pairs per pulse can become significant. This has
the effect of reducing the Interference Visibility and hence at a
more fundamental level the entanglement. This is a critical point
especially in experiments of this nature. We ensured that we had
the same mean power at the PPLN waveguide throughout these
experiments. This was done so that any reduction in the
Visibilility was due to decoherence and not due to variation in
the rate of production of photon pairs.

A theory describing this behaviour has recently been developed,
the details of which will be presented elsewhere
\cite{Marcikic02a}. From this we expect the Visibility to be
related inversely to the number of photon pairs produced, $V=1/n$.
If we also make the assumption that the pairs satisfy a Poissonian
distribution, where $P(n) =\mu^ne^{-\mu}/n!$ is the probability of
having $n$ pairs and $\mu $ is the mean number of photon pairs
created per pulse, then we can find the following relationship
\begin{eqnarray}
V(\mu) = V_{max}\frac{e^{-\mu}}{1 - e^{-\mu}}
\sum_{n\ge1}\frac{\mu^n}{n!n}.
\end{eqnarray}
Here $V_{max}$ is a scaling factor to compensate for known factors
that reduce the Visibility in a given experiment. From our
knowledge of the experimental setup we can relate the measured
quantities, the 2 single photon count rates, $S_1, S_2$, and the
average coincidence count rate, $R_c$, to give us an estimate of
the mean number of photon pairs, $ \mu_e = S_1 S_2/4 R_c f$, where
$f$ is the frequency of the pulsed laser source. This gives us a
measure that not only takes account of collection and detection
efficiencies but also losses, which is convenient when working
over long distances. We see in Fig.(\ref{fig:pairvis}) that theory
and experiment are in good agreement which allowed us to optimise
the power so as to efficiently generate arbitrarily entangled
states for transmission.
\begin{figure}[h]
\epsfig{figure=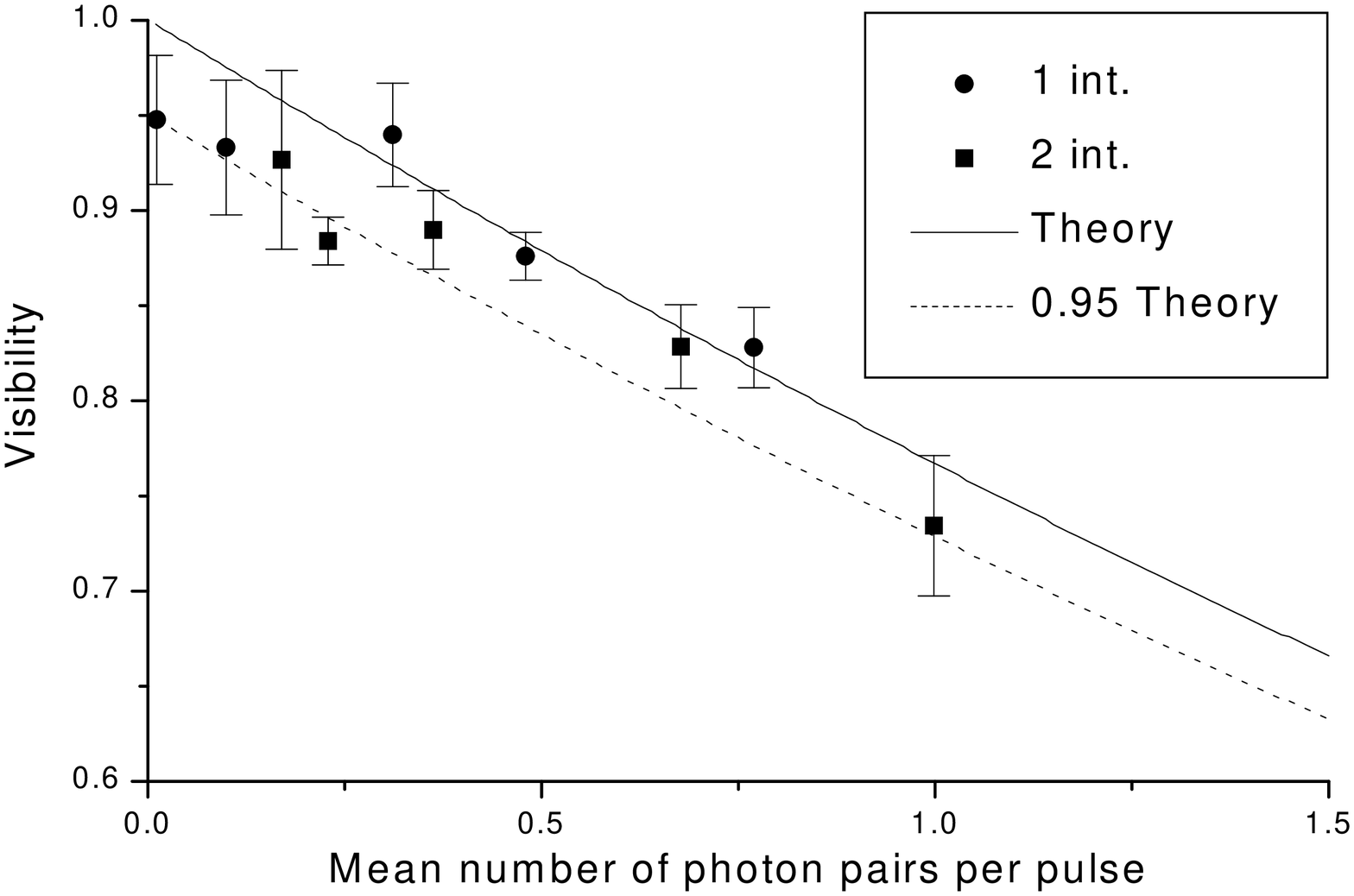,width=80mm,height=48mm} \caption{
Interference Visibility as a function of the mean number of Photon
Pairs in each pulse of the laser. \label{fig:pairvis} }
\end{figure}

We have shown that we can generate, in a controlled way, states
with arbitrary degrees of entanglement using a pulsed diode laser
and a PPLN waveguide. This is very much a new technology under
development which holds great promise for integrated quantum
optics and hence quantum communication. The main results proved
resoundingly that Time-Bin entanglement is robust with
transmission distances up to 11\,km.

This work was supported by the  Swiss NCCR "Quantum Photonics" and
the European QuComm IST projects. W.T. acknowledges funding by the
ESF Programme Quantum Information Theory and Quantum Computation
(QIT)

\end{document}